\documentclass[12pt,preprint]{aastex}
\usepackage{color}
\usepackage{lscape}
\usepackage{epsfig}
\defcitealias{jeffery93a}{JH93}

\slugcomment{Draft version August 2021.}

\shorttitle{Revised abundances of RCB stars}
\shortauthors{Pandey, Hema, Reddy}

\begin{document}

\title{Revised surface abundances of R Coronae Borealis stars}

\author{Gajendra Pandey, B. P. Hema, and Arumalla B. S. Reddy}
\affil{Indian Institute of Astrophysics,
Bangalore, Karnataka, 560034, India}
\email{pandey@iiap.res.in; hemabp.phy@gmail.com; balasudhakara.reddy@iiap.res.in}

\begin{abstract}

Surface abundances of 14 (11 majority class and 3 minority class) R\,Coronae Borealis stars (RCBs) along with 
the final flash object, V4334\,Sgr (Sakurai's object) are revised based on their
carbon abundances measured from the observed C$_2$ bands; note that the earlier reported abundances were derived using
an assumed carbon abundance due to the well known ``carbon problem''. The hot RCB MV\,Sgr is not subject to a carbon problem;
it is remarkable to note that MV\,Sgr's carbon abundance lies in the range that is measured for the majority and minority class RCBs.
The revised iron abundances for the RCBs are in the range $\log\epsilon$(Fe)$=$3.8 to $\log\epsilon$(Fe)$=$5.8 with the minority class 
RCB V854\,Cen at lower end and the majority class RCB R\,CrB at the higher end of this range. Indications are that the revised 
RCBs' metallicity range is roughly consistent with the metal poor population contained within the bulge. The revised abundances of RCBs 
are then compared with extreme helium stars (EHes), the hotter relatives of RCBs. Clear differences are observed between RCBs and EHes 
in their metallicity distribution, carbon abundances, and the abundance trends observed for the key elements. These abundances are further
discussed in the light of their formation scenarios.

\end{abstract}

\clearpage

\section{Introduction}

R\,Coronae Borealis stars (RCBs) are F- and G-type supergiants with 
carbon rich and hydrogen deficient atmospheres. In these stars, the measured hydrogen-to-helium ratio 
by number is $\le$ 10$^{-2}$ that stands in contrast to a solar-type star's H/He of 10. The surface abundances of
hydrogen poor stars are measured relative to helium, the most abundant element in their atmospheres.
The next most abundant element in their atmospheres is carbon followed by nitrogen and oxygen, and the 
rest are the trace elements including hydrogen. Spectroscopic determination of carbon-to-helium ratio is not 
possible from the observed optical spectra of RCBs $-$ photospheric neutral helium lines are not expected for 
their effective temperatures and neutral carbon lines are roughly of the same strength across the range in 
their effective temperature unlike the lines from other elements, for example, iron. Hence, a carbon-to-helium 
ratio of 1.0\% by number is assumed to derive the surface abundaces of RCBs. This assumption comes from the 
extreme helium stars (EHes), seemingly the close relatives of RCBs, having C/He ratios of about 1.0\%.

Of about 128 known Galactic RCBs \citep{tisserand2020} including the 11 new RCBs from 
Palomar Gattini IR (PGIR) survey \citep{karambelkar2021}, abundance analyses are available for only
about 22 $-$ a very small sample indeed \citep{asplund2000,raonlambert2008,hema2017} to draw conclusions on their
origin and evolution as a group.

In this paper, the dependence of the derived surface abundances on the 
adopted model atmosphere's C/He ratios are discussed. Here, we report the revised surface abundances 
based on the derived carbon abundance, that is the C/He ratio, from the observed C$_2$ bands in an RCB's 
spectrum \citep{hema2012,hema2017}. These new abundances are further compared with the abundances of EHes, and are 
then discussed in the light of their formation scenarios.

\section{Abundances}

\subsection{Fundamentals}

In an observed star's spectrum, the strength of an absorption line is always defined as the depth of the line
with respect to the continnum. Thus, the strength of an absorption line (weak) of an element X is controlled 
by the line absorption coefficient relative to the continuum absorption coefficient. In the case of a 
normal (H-rich) star's optical spectrum with effective temperature in the range similar to RCBs, the continuum 
and the line absorption coefficients are proportional to the number density of H and X atoms, respectively.
Then, the number density ratio X/H is what dictates the line strength and is normally expressed as the 
abundance of an element X. X/H can be further stated in terms of the mass fraction Z(X) by assuming 
He/H ratio of 0.1 that comes from hot normal stars and is an unobservable quantity for cool stars.

In the optical, theory and observations of RCBs' atmospheres suggest that photoionization of C\,{\sc i} is the dominant source of
continuum opacity in their line-forming layers \citep{asplund1997a}. The near constant equivalent widths of weak C\,{\sc i} lines
observed in their optical spectra from one star to another is notable (see Figure 1 of \citet{raonlambert96}).
Therefore, the measure of abundance of an element X in RCBs is the ratio X/C. However, it is crucial to express 
the measured abundance A(X) = X/C in terms of the more fundamental quantity: the mass fraction Z(X). Which calls
for the determination or assumption of the C/He ratio, since helium is expected to be the most abundant element in their
atmospheres. The mass fraction, Z(X) is given as

\begin{equation}
Z(X)=\frac{\mu_X X}{\mu_H H + \mu_{He} He + \mu_C C + ...}
= \frac{\mu_X X}{(\sum\mu_I I)}
\end{equation}

where $\mu_{I}$ is the atomic mass of element I. The denominator which represents summation over all the elements is a
conserved quantity through the different stages of nuclear burning. Assuming helium to be the most abundant ingredient,
the equation 1 in terms of the measured abundance A(X) = X/C is redefined as

\begin{equation}
Z(X)= \frac{\mu_X A(X)}{H/C + 4He/C + 12 + ..\mu_I I/C }
\end{equation}

Due to hydrogen being very poor in these stars, H/C relative to He/C is very very small and can be ignored like other 
trace elements from the denominator, then the above equation reduces to,

\begin{equation}
Z(X) \simeq \frac{\mu_X}{4}\frac{C}{He}A(X)
\end{equation}

The C/He is spectroscopically determined for 14 RCBs from their observed C$_2$ bands \citep{hema2012,hema2017}, and also the abundance 
of any element X for a hydrogen-deficient star like RCBs, can be directly measured spectroscopically i.e., A(X) = X/C.

The derived abundances are normalised based on the convention that $\log\epsilon$(X) = $\log(X/H)$ + 12.0 to a scale 
in which $\log\sum\mu_I\epsilon(I)$ = 12.15, where 12.15 is determined from solar abundances with  He/H $\simeq$ 0.1. 
Based on this normalisation convention, from equation 3, the helium abundance $\log\epsilon$(He) is about 11.54 for 
a C/He $\le$ 0.01. The abundance of an element X can be expressed in terms of mass fraction Z(X) or number fraction X/He, and these
two quantities are related as shown in equation 3. Here, stars' derived abundances are given in $\log\epsilon$(X) and the notation [X] 
represents abundance of X in a star relative to that in the Sun or log number relative to solar. 

\subsection{RCB stars}

The surface abundances of about 22 RCBs are available in literature as mentioned in Section\,1. These abundances
were derived by adopting a C/He of 1\%, or that is by assuming the carbon abundance $\log\epsilon$(C) = 9.54.
Note that, from equation 3, the derived abundances depend on the adopted C/He ratio but the abundance ratios do not.
For example, $\log\epsilon$(X/Fe) is independent of the adopted C/He ratio. Four RCBs: VZ\,Sgr, V\,CrA, V854\,Cen, and V3795\,Sgr,
are clasified as minority class RCBs that show relatively lower Fe abundances and higher Si/Fe and S/Fe ratios than the
majority class RCBs \citep{asplund2000}.

In this paper, the surface abundances of 14 RCBs (11 majority class and 3 minority class) including their fluorine abundances 
\citep{pandeyetal2008,hema2017} are revised using their spectroscopically determined 
carbon abundances from observed C$_2$ bands \citep{hema2012,hema2017}. The revised abundances for all the elements
except helium, can be obtained by a simple scaling down of the derived abundances (for an assumed C/He) by a factor, 
that is the difference between the assumed and the determined $\log\epsilon$(C). The factor, $\Delta$C, by which the 
derived abundance (for an assumed C/He) was scaled down is given in Table 1. Before presenting their revised abundances, 
we comment on the two known facts related to their derived carbon abundances.

First, the predicted strengths of the C\,{\sc i} lines are stronger than the observed. That means, the carbon abundance
derived using state-of-the-art H-deficient model atmospheres \citep{asplund1997a} is about 0.6 dex reduced from that
chosen for costructing the model atmosphere. \citet{asplund2000} gave a name the ``carbon problem'' to this mismatch.
A more severe ``carbon problem'' was found for the observed [C\,{\sc i}] lines \citep{pandey2004b}. Resolutions to the 
``carbon problem'' were provided by crafting state-of-the-art model atmospheres by hand \citep{asplund2000,pandey2004b}. 
\citet{pandey2004b}'s hand crafted model atmospheres bring the predicted strengths of C\,{\sc i} and [C\,{\sc i}] lines
in agreement with the observations. However, this holds for all the carbon abundances, from 
$\log\epsilon$(C)$=$8.5 (C/He$=$0.1\%) through $\log\epsilon$(C)$=$10.5 (C/He$=$10\%), adopted for constructing the 
model atmosphere. This implies that predicted strengths of C\,{\sc i} and [C\,{\sc i}] lines are insensitive to the 
carbon abundance. 

Second, the carbon abundances derived from observed C$_2$ bands are independent of the adopted model's carbon abundance 
and this is described as the C$_2$ carbon problem \citep{hema2012}. The solution to the so called C$_2$ carbon problem
may lie in the modification of the model atmosphere's temperature structure as shown by \citet{pandey2004b}. However, it is 
yet to be shown that real atmospheres have flatter temperature gradients, as suggested by \citet{asplund2000}, than the 
present state-of-the-art model atmospheres. Therefore, it cannot be ruled out that the carbon abundances derived from 
C$_2$ bands are the real measure of carbon abundances in these stars. In principle, the carbon abundances measured from 
C$_2$ Swan bands and that assumed for the model atmosphere can be equated for a particular choice of C/He that varies from 
star to star. This removes the carbon problem for C$_2$ bands. Also, as expected for carbon rich stars, the carbon abundance derived
from C$_2$ bands is correlated with the O abundance \citep{hema2012}. (Note that the carbon abundance from C\,{\sc i} lines is not 
well correlated with the O abundance). Hence, we adopt the carbon abundance or the C/He ratio derived from 
C$_2$ bands \citep{hema2012,hema2017} for revising the RCBs' abundances. The revised abundances, deduced from the above described 
procedure, are given in Table 1.

\section{The Galactic positions and orbits}

It is crucial to know whether RCBs and their apparent relatives, the EHes are members of the Milky Way's bulge (the central regions) 
or the halo (the outer regions). To confirm their membership, the distances and orbits of these stars have been
determined. The distances to these stars were determined using the Parallax measurements made by the Gaia satellite.
Note that, the Gaia mission measures a star's proper motion and radial velocity (RV) with unprecedented 
precision. The RV measurements of several RCBs and EHes also come from \citet{asplund2000,hema2012,pandeyetal2001,jefferyetal1987,pandey1996} 
and the best or an average value is adopted based on our judgement. The orbits around the Galaxy have been estimated in combination
with their distances and velocities. The required data for calculating the orbits for the following RCBs: GU\,Sgr, UX\,Ant, R\,CrB, 
RS\,Tel, SU\,Tau, V482\,Cyg, FH\,Sct, V2552\,Oph, V532\,Oph, ASAS-RCB-10, VZ\,Sgr, and MV\,Sgr (the hot RCB) were available. For the
following EHes: LSS\,3184, BD$-$9$^{\circ}$4395, LS\,IV+$6^{\circ}$002, LSE\,78, LSS\,4357, V1920\,Cyg, LSS\,99, HD\,124448, 
BD+10$^{\circ}$2179, PV\,Tel, FQ\,Aqr, LS\,IV$-$1$^{\circ}$ 002, BD$-$1$^{\circ}$3438, LS\,IV$-$14$^{\circ}$ 109, and LSS\,3378,
data were available for calculating the orbits.
The initial conditions for the computation of Galactic orbits of stars are their presently observed positions and 
velocities with respect to the galactocentric reference frame. Adopting the 
solar motion (U, V, W)$_{\odot}$ = (10.0, 5.2, 7.2) km s$^{-1}$ from \citet{dehnenbinney1998}, the local standard of 
rest (LSR) velocities of stars U$_{LSR}$, V$_{LSR}$, W$_{LSR}$ and their errors $\sigma_{U}$, $\sigma_{V}$, $\sigma_{W}$ 
are calculated with the method of \citet{johnsonnsoderblom1987}. The LSR velocities are then corrected to the 
Galactic standard of rest (GSR) by adopting the LSR rotation velocity of 220 km s$^{-1}$ \citep{kerrnlynden1986} at 
the galactocentric distance of the Sun of 8.5 kpc. The celestial positions ($\alpha$, $\delta$, l, b), parallaxes ($\pi$), 
and absolute proper motions ($\mu_{\alpha}$\,cos $\delta$, $\mu_{\delta}$) are adopted from the 
Gaia Early Data Release 3 \citep{gaiacollab2016b,gaiacollab2020a}, while the RV measurements are taken from the published
sources as mentioned above. Adopting the measured Galactic spatial positions and velocities, we studied the 
dynamics of stars under the influence of a multicomponent, static, axisymmetric Galactic gravitational potential model 
of \citet{flynnetal1996}. The relevant code adopted here for the integration of stellar orbits was used previously in the 
analysis of kinematics and orbits of a large sample of open clusters \citep{wuetal2009,reddyetal2016}. Starting with a 
star's current position and velocity components referenced to the Galactic standard of rest, the trajectory of the star was 
followed backward in time over a period of 5 Gyr to ensure that each star could complete sufficient galactic orbits so that 
the averaged orbital parameters can be determined with fair certainty.

The analysed RCBs with accurate kinematics have tightly bound orbits, placing them in the inner regions of the Milky Way of radius less
than or about 6.0 kpc and z$_{max}$ about 3.0 kpc, the exceptions being UX\,Ant, V482\,Cyg, SU\,Tau, and R\,CrB. In contrast, the analysed EHes
with accurate kinematics have orbits extending beyond the regions of the Milky Way of radius more than 6.0 kpc, the exceptions being
LSS\,4357, LS\,IV$-$1$^{\circ}$ 002, and LS\,IV$-$14$^{\circ}$ 109. Including MV\,Sgr, we note that about five analysed RCBs
with accurate kinematics have tightly bound orbits and z$_{max}$ similar to the most metal-poor, $\log\epsilon$(Fe)$\sim$3.5, 
star SMSS\,J181609.62$-$333218.7 whose orbit is entirely contained within the bulge \citep{howesetal2015}. The revised metallicity range for 
the analysed RCBs is $\log\epsilon$(Fe)=3.8 through $\log\epsilon$(Fe)=5.8 (see Table 1), and the revised metallicities are fairly consistent 
with their estimated orbits and location in the Galaxy.

\section{RCB and EHe stars: surface compositions}

The RCBs' revised surface compostion, based on the carbon abundances derived from the observed C$_2$ bands, provide new evidences
for their formation history. We discuss by comparing the revised surface compositions of RCBs with EHes that are considered to be their
relatives having higher effective temperatures. The surface compostion of EHes including the hot RCB star DY\,Cen are from 
\citet{jefferynheb1992,drillingetal1998,jeffery1998,jefferyetal1998,pandeyetal2001,pandeyetal2004a,pandeyetal2006,pandey2006,pandeynreddy2006,pandeynlamb2011,pandeyetal2014,pandeynlamb2017,jeffery2017,bhowmicketal2020}. 
The principal objective is to seek out similarities, differences, and trends if any. 

\subsection{The C/He ratios}

The carbon abundances for the RCBs are in the range $\log\epsilon$(C)$=$7.7 to $\log\epsilon$(C)$=$8.8 that is the C/He ratios
are in the range 0.02 to 0.2 per cent including the three minority class RCB stars. In contrast, the C/He ratios of the EHes are in the range 
0.3 to 1.0 per cent with three exceptions. The three exceptions are V652\,Her, HD\,144941, and GALEX J184559.8$-$413827, having very low
C/He ratios $\sim$ 0.003 per cent. The two hot RCBs, MV\,Sgr \citep{jeffery1988} and DY\,Cen have C/He ratios of 0.02 and 1.0 per cent,
respectively. However, recent study of \citet{jeffery2020} indicates that DY\,Cen's evolutionary history involves a very late thermal
pulse due to its very rapid evolution, non-negligible surface hydrogen and high surface strontium. 

The hot RCB MV\,Sgr is not subject to a carbon problem like EHes. It is remarkable to note that MV\,Sgr's C/He ratio lies in the range that 
is derived for the majority and minority class RCBs.

\subsection{Iron: the initial metallicity}

The revised iron abundances for the RCBs are in the range $\log\epsilon$(Fe)$=$3.8 to $\log\epsilon$(Fe)$=$5.8 with the minority
class RCB V854\,Cen at lower end and the majority class RCB R\,CrB at the higher end of this range (see Table 1). On the contrary,
the iron abundances for the EHes are in the range $\log\epsilon$(Fe)$=$5.4 to $\log\epsilon$(Fe)$=$7.2. As discussed in Section 3,
EHes are broadly placed in the outer regions of the Milky Way than the RCBs. Indications are that the RCBs' metallicity range is roughly 
consistent with the metal poor population contained within the bulge \citep{howesetal2015}.

Hydrogen- and helium-burning products are clearly observed in the atmospheres of RCBs and EHes. Their derived effective temperatures
and surface gravities suggest that they are evolved low mass stars. Hence, no synthesis of $\alpha$- and Fe-peak elements occurs in 
the course of their evolution. The $\alpha$- and Fe-peak elements remain unaltered in their atmospheres by providing us the initial
metallicty to these stars; here $\alpha$-peak elements usually refer to $\alpha$-capture elements heavier than neon. 

\subsection{Hydrogen}

The hydrogen abundance of majority class RCB stars show no obvious trend with the iron abundance or the metal abundance. However, it is 
notable that two minority class RCBs V\,CrA and V854\,Cen with relatively lower Fe abundance have hydrogen abundance higher than the 
majority class RCBs. The third minority class RCB VZ\,Sgr is like the majority class RCBs (See Figure 1: bottom left panel). For a comparison,
EHe stars are shown in Figure 1: bottom right panel. Note the relatively higher hydrogen abundances of DY\,Cen and the three very low C/He 
EHe stars.

\subsection{The CNO abundances}

The majority class RCBs' carbon abundances are a function of their Fe abundances but this trend is not very evident for the minority
class RCBs (see Figure 1: top left panel). However, abundances for only three minority class RCBs are available and so it is not 
worth looking for any trends. In contrast for EHes, Figure 1: top right panel clearly demonstrates that carbon abundances are
independent of their Fe abundances. 

The nitrogen abundances of both RCBs and EHes depend on their Fe abundances (see Figure 2: bottom left panel for RCBs and bottom right panel 
for EHes). The observed N abundance is the sum of initial CNO as expected.  Hence, providing the evidence that helium is produced from 
hydrogen-burning CNO cycle that converts most of the initial C and O to N. 

If all of initial C and O is coverted to N, then the CNO cycle processed material should be N enriched with C and O depleted. The observed
C and O abundances for both RCBs and EHes are not depeleted suggesting that these are products of helium-burning via triple-$\alpha$ and
$\alpha$-capture on $^{14}$N and $^{12}$C. The three exceptions that show depleted C and O are the very low C/He or C-poor EHe stars 
(see Figure 1: top left panel for RCBs and top right panel for EHes, and Figure 2: top left panel for RCBs and top right panel for EHes).

\subsection{Fluorine}

The fluorine abundances of the majority and minority class RCBs suggest a mild trend with their Fe abundances and F/Fe in these stars is highly 
enriched when compared to the solar F/Fe ratio. Most of the EHes have enhanced F like RCBs but show no trend with their Fe abundances unlike RCBs. 
See Figure 3: bottom left panel for RCBs and bottom right panel for EHes, for F versus Fe trends.

\subsection{Neon}

For EHes, the neon abundances like N abundances are enhanced and show a trend with their Fe abundances except for the five cool EHes whose Ne 
abundances are not corrected for the non-LTE effects. Application of the non-LTE effects brings the Ne abundances of cool EHes in line with the
hot EHes \citep{pandeynlamb2011}. The dependence of the Ne abundance with the Fe abundance suggest that the observed Ne in EHes is essentially 
$^{22}$Ne and is produced from two successive $\alpha$-captures on $^{14}$N. Note that the revised Ne abundances are available only for two
RCBs. The Ne versus Fe trends are shown in Figure 3: top left panel for RCBs and top right panel for EHes.

\subsection{Sodium to zinc}

The EHe and RCB abundances of sodium, aluminium, magnesium, silicon, sulphur, calcium, titanium, chromium, manganese, nickel, and zinc scale well 
with the iron abundances. For RCBs, Al, Mg, Si, S, Ca, Ti, Ni and Zn broadly vary in concert with Fe, as expected. Similar correlations are seen for 
EHes with additions of Cr and Mn. This clearly indicates that the EHe as well as RCB star's Fe abundance is the representative of initial metallicity. 
Abundances of phosphorus, argon, chromium, and manganese are not available for RCBs, but in EHes it is worth considering the observed abundances of P 
and Ar that suggest weak correlations with Fe. The X versus Fe trends are shown in Figures 4, 5, 6, 7, 8, 9, and 10. The left and the right panels 
show RCBs and EHes, respectively. Note the variation of minority class RCBs with respect to the majority class RCBs.

\subsection{Yttrium, zirconium, and barium: heavy elements}

For majority class RCBs, the revised abundances of Y and Zr show insignificant enhancements. However, the minority class RCBs show a range from 
insignificant to a maximum enhancement of about 2.0 dex in both Y/Fe and Zr/Fe with respect to the solar Y/Fe and Zr/Fe ratios like the EHes. 
Barium for majority as well as minority class RCBs shows insignificant enhancement. Note that only three measurements of Ba are available for EHes 
showing a range from no to a maximum enhancement of about 1.8 dex in Ba/Fe with respect to the solar Ba/Fe ratio. See Figures 10 and 11: the left 
and the right panels show RCBs and EHes, respectively.

\section{Key spectroscopic features}

The key features of RCBs are specifically their high $^{12}$C/$^{13}$C and low $^{16}$O/$^{18}$O ratios with remarkable F overabundances \citep{hema2012,
clayton07,pandeyetal2008}. $^{12}$C/$^{13}$C and $^{16}$O/$^{18}$O ratios are from observed C$_2$ and CO molecular bands in the spectra of RCBs but these
molecular bands are not present in the observed spectra of EHes due to their higher effective temperatures. Nevertheless, fluorine atomic lines are
observed in EHes' as well in RCBs' spectra and provide the star's F abundance. The atmospheres of EHes are overabundant in F like RCBs \citep{pandey2006,
bhowmicketal2020}. The processes involving fluorine production in these stars needs to be explored. For this reason, it will be crucial 
to identify any correlations between fluorine and other elements including any relationship among the abundances of other key elements. 

\subsection{F versus CNO and Ne}

In RCBs, the F abundances suggest mild to no correlation with C abundances. In contrast, EHes' F abundances are strongly correlated with their C 
abundances. See Figure 12: the top left and the top right panels show RCBs and EHes, respectively.

F versus N show significant correlation in RCBs unlike in EHes. Figure 13: the top left and the top right panels clearly exhibit these
trends for RCBs and EHes, respectively. F versus O and Ne trends are also shown in Figure 14: the left and the right panels 
are for RCBs and EHes, respectively.

We note that the relationship of F with C and that with N is distinct for RCB and EHe stars. This indicates that, possibly, fluorine is produced from
two different processes operating in these stars.

\subsection{C and N versus O}

C and O abundances suggest linear correlation for both RCBs and EHes. These trends are clearly shown in Figure 12: the bottom left and the bottom right
panels are for RCBs and EHes, respectively. The N versus O abundances are also shown in Figure 13: the bottom left and the bottom right 
panels are for RCBs and EHes, respectively.

\section{Double white dwarf mergers and the abundances of key elements}

The RCB and EHe stars' origins and evolutionary connections are not yet understood despite thorough analyses of their spectra. In broad terms, 
the chemical compositions suggest a hydrogen-deficient atmosphere now composed of material exposed to both H- and He-burning. Following the elimination 
of several proposals, two principal theories emerged: the ``double-degenerate'' (DD) model and the ``final-flash'' (FF) model. 

The FF model, refers to a late or final He shell flash in a post-AGB star. In this model \citep{iben1983}, the ignition of the helium shell in a 
post-AGB star, say, a cooling white dwarf, results in what is known as a late or very late thermal pulse \citep{herwig2001}. This converts the star 
to a H-poor cool luminous star (i.e., an RCB star), which then evolves to hotter effective temperatures at about constant luminosity (i.e., as an EHe star), 
and finally to the white dwarf cooling track. The attendant nucleosynthesis during and following the He shell flash shows that a H-poor supergiant may
result with surface composition characteristic of RCBs and EHes. However, the  FF model has failed to account for the key features, particularly,
the high $^{12}$C/$^{13}$C ratios, the low $^{16}$O/$^{18}$O ratios and anomalous F overabundances observed in these stars \citep{pandey2006,clayton07,
pandeyetal2008,raonlambert2008,hema2012,hema2017}. A consensus is now emerging for DD model but a small fraction of H-poor stars may be produced from 
FF model \citep{pandeynlamb2011} such as the majority RCB XX\,Cam, and possibly the EHe HD\,124448 \citep{bhowmicketal2020} including
V4334\,Sgr (Sakurai's object) \citep{pandeyetal2008}. The surface abundances of V4334\,Sgr \citep{asplund97b} are revised based on the measured carbon
abundance, that is $\log\epsilon$(C) = 9.7 \citep{hema2012b}, from  C$_2$ bands as done for RCBs. For comparison, the revised abundances of V4334\,Sgr 
for the 1996 October spectrum are given in Table 1 .

The DD scenario, proposed by \citet{webbink84} and \citet{iben84}, involves the merger of an He white dwarf with a more massive C-O white dwarf 
following the decay of their orbit. Other mergers may involve two He white dwarfs. It is clear that the merger product will be H-poor since neither 
of the white dwarfs contain much hydrogen, and the hydrogen that survives will be mixed substantially with more helium and possibly other material.
Recall the extraordinary $^{16}$O/$^{18}$O ratios in RCBs and/or the remarkable F overabundances in RCBs as well as in EHes. Neither CO+He 
white dwarf binaries nor He+He white dwarf binaries can account for these exceptional abundances without the ensuing nucleosyntheis during
the merger and/or the postmerger phase. Simulations of the merger and postmerger phases with accompanying nucleosynthsis have been attempted for 
evolution of a white dwarf merger to the RCB phase \citep{longland11,zhang12a,zhang12b,menon13,menonetal2019,zhangetal2014,lauer2019}.

Simulations of He+He white dwarf mergers are limited to \citet{zhang12a,zhang12b} and appear to be ``in partial agreement" in explaining the observed 
abundances of RCBs and EHes. \citet{bhowmicketal2020} note that F is underpredicted by \citet{zhang12b} while minor disagreements between prediction
and observation are found for C, N, O, and Ne. In these mergers, the F is synthesized by 
$^{14}$N$(\alpha,\gamma)^{18}$F$(p,\alpha)^{15}$O$(\alpha,\gamma)^{19}$Ne$(\beta^+)^{19}$F. 
However, \citet{zhang12a} provide the latitude to account for the C-poor V652\,Her and HD\,144941 \citep{pandeynlamb2017}.

Most recent calculations for a CO+He merger are from \citet{crawford2020}. It is clear that surface composition of the merger products are
a result of mixing during and/or following the merger. In post-merger objects, the temperature of the helium-burning shell at ignition strongly impacts 
the yields of CNO-process as well as $\alpha$-capture isotopes. Hence, Crawford et al. investigate the effects of a range of initial He-burning shell 
temperatures and include the effects of solar and subsolar metallicities. Their three models that satisfy the maximum criteria in reproducing the 
observed surface abundances are SOL8.57, SUB8.48, and SUB8.53; SOL and SUB represent models with metallicity solar and one-tenth of solar, respectively, 
and 8.57, 8.48, and 8.53 are the models' initial He-burning shell temperatures (K) in common log values. Crawford et al. were able to identify SUB8.48 
as the preferred model that is able to remarkably reproduce abundances closest to those of observed RCBs. SUB8.48 is at 10 per cent of solar
metallicity with an initial He-burning shell temperature of approximately 3.00$\times$10$^8$ K; the other closest model SUB8.53 yields higher C, F, and
Ne with no significant changes in N and O yields (see Figures 1, 2, and 3). The predicted surface Li abundance ($\log\epsilon$(Li)) is between 0.85 and 
2.64 for solar models, while the subsolar models have significantly higher Li abundances, between 3.95 and 6.41. 
The Crawford et al.'s predicted surface Li can possibly be an upper limit because their reaction network, mesa\_75.net, as noted by them does not include
the crucial reaction $^{7}$Li$(\alpha,\gamma)^{11}$B. However, inclusion of $^{11}$B in the reaction network reduces the predicted surface Li 
abundance ($\log\epsilon$(Li)) significantly to about $-$1.5 to 1.0 for one-tenth of solar metallicity white dwarf mergers \citep{munson2021}. 

\section{Concluding Remarks}

Predictions of CO+He white dwarf mergers are in good agreement with the observed abundances, in particular with the extraordinary overabundance of F
\citep{menon13,menonetal2019}. \citet{menon13} identify that one source of F is in the He-burning shell where $^{13}$C$(\alpha,n)^{16}$O provides
the neutrons to seed the reaction $^{14}$N$(n,p)^{14}$C$(p,\gamma)^{15}$N$(\alpha,\gamma)^{19}$F. The correlations of F with C and N indicate that
plausibly F production in majority class RCBs is via the reaction channel:
$^{14}$N$(\alpha,\gamma)^{18}$F$(p,\alpha)^{15}$O$(\alpha,\gamma)^{19}$Ne$(\beta^+)^{19}$F; F abundance depends on $^{14}$N and the available
protons (see Section 5.1, and Figures 12 and 13).
While in EHes, the dependence of F with C and N suggests that F is produced via the reaction channel:
$^{14}$N$(n,p)^{14}$C$(p,\gamma)^{15}$N$(\alpha,\gamma)^{19}$F; F abundance depends on the available neutrons but on the other hand, $^{13}$C which
provides neutrons must result from the mixing of protons into a $^{12}$C-rich region, where the $^{12}$C comes from $^{4}$He
(see Section 5.1 and Figures 12 and 13). For example, if F synthesis is via the reaction channel involving neutrons, then the star's F abundance is
expected to be correlated with its $^{12}$C abundance. On the contrary, if the reaction involving $\alpha$-capture on $^{14}$N is producing F, then
the star's F abundance is expected to be correlated with its N abundance. Nonetheless, the trends of F with C, N, O, and Ne show that significant
helium burning after a double white dwarf merger can account for a majority of the observed abundances (see Section 5.1, and Figures 12, 13, and 14).
It is worth noting that the average of observed F abundances is lower by about 1 dex in RCBs than in EHes for stars with common
Fe abundance (see Figure 3).

The revised Fe abundances ($\log\epsilon$(Fe)), that is the metallicity, of RCBs are in the range 3.8 through 5.8. These revised metallicities 
are fairly consistent with their estimated orbits and location in the Galaxy. Recent \citep{crawford2020,munson2021} 
and the earlier predictions of CO+He white dwarf mergers do not explore the metallicity range lower than $\log\epsilon$(Fe)$=$6.5.
The revised metallicity range observed for RCB stars is lower than $\log\epsilon$(Fe)$=$6.5. Hence, predictions for CO+He white dwarf mergers 
are lacking for the revised lower metallicities of RCB stars. However, an extrapolation of Crawford et al.'s
surface abundance  preditions to lower metallicities, similar to the revised metallicity range of RCBs, would likely explain the observed carbon 
abundances from C$_2$ bands but not all the observed key elemental abundances. The revised C/O ratios that are between 3.0 and 64.0 may possibly 
be explained; tweaking of several parameters mainly the initial He-burning shell temperature may likely reproduce the RCBs' revised abundances. 
Theoretical studies with a larger parameter space, for example including all the key reactions, needs to be explored to explain the revised 
abundances of RCBs, and this would certainly lead to further refinements in the study of white dwarf mergers.

\acknowledgments
We thank the referee for the constructive comments. GP is grateful to Kameswara Rao, David Lambert and Simon Jeffery for introducing him to the topic 
of H-poor stars, and for all their help.

\begin{figure}
\epsscale{1.00}
\plotone{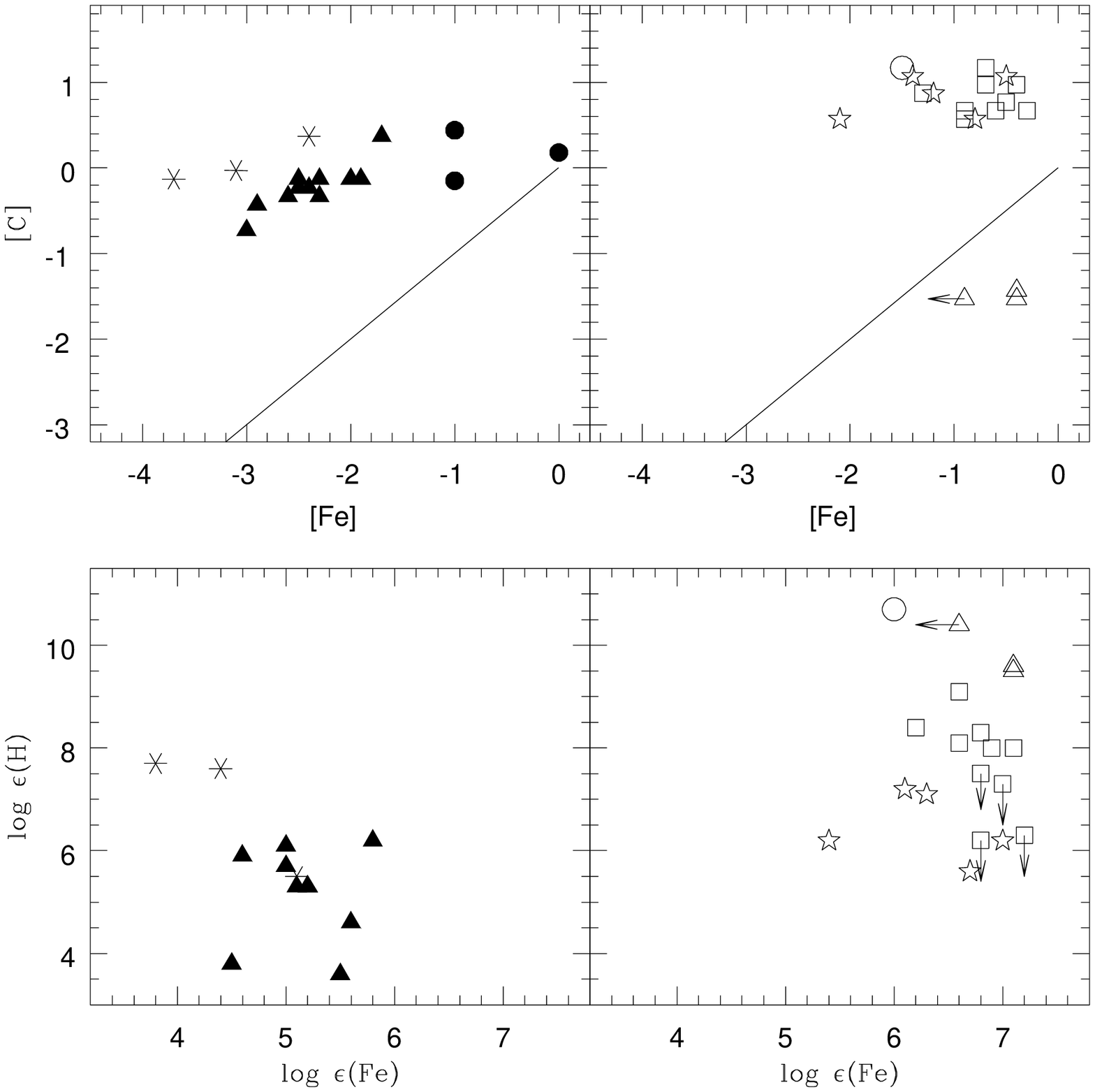}
\caption{$\log\epsilon$(H) vs. $\log\epsilon$(Fe) for RCBs (bottom left panel) and EHes (bottom right panel). 
[C] vs. [Fe] for RCBs (top left panel) and EHes (top right panel). The majority and minority class RCBs are 
represented by filled triangles and asterisks, respectively. The hot and cool EHes are represented by open squares
and open stars, respectively. C-poor EHes are represented by open triangles. DY\,Cen is represented by open circle.
Filled circles denote the model predictions from \citet{crawford2020} for three Model IDs SOL8.57, SUB8.48, 
and SUB8.53, meeting the maximum criteria; details are in the text of Section 6.
[C]$=$[Fe] is denoted by solid line, and [Fe],[C]$=$0,0 represents the Sun.}
\end{figure}

\begin{figure}
\epsscale{1.00}
\plotone{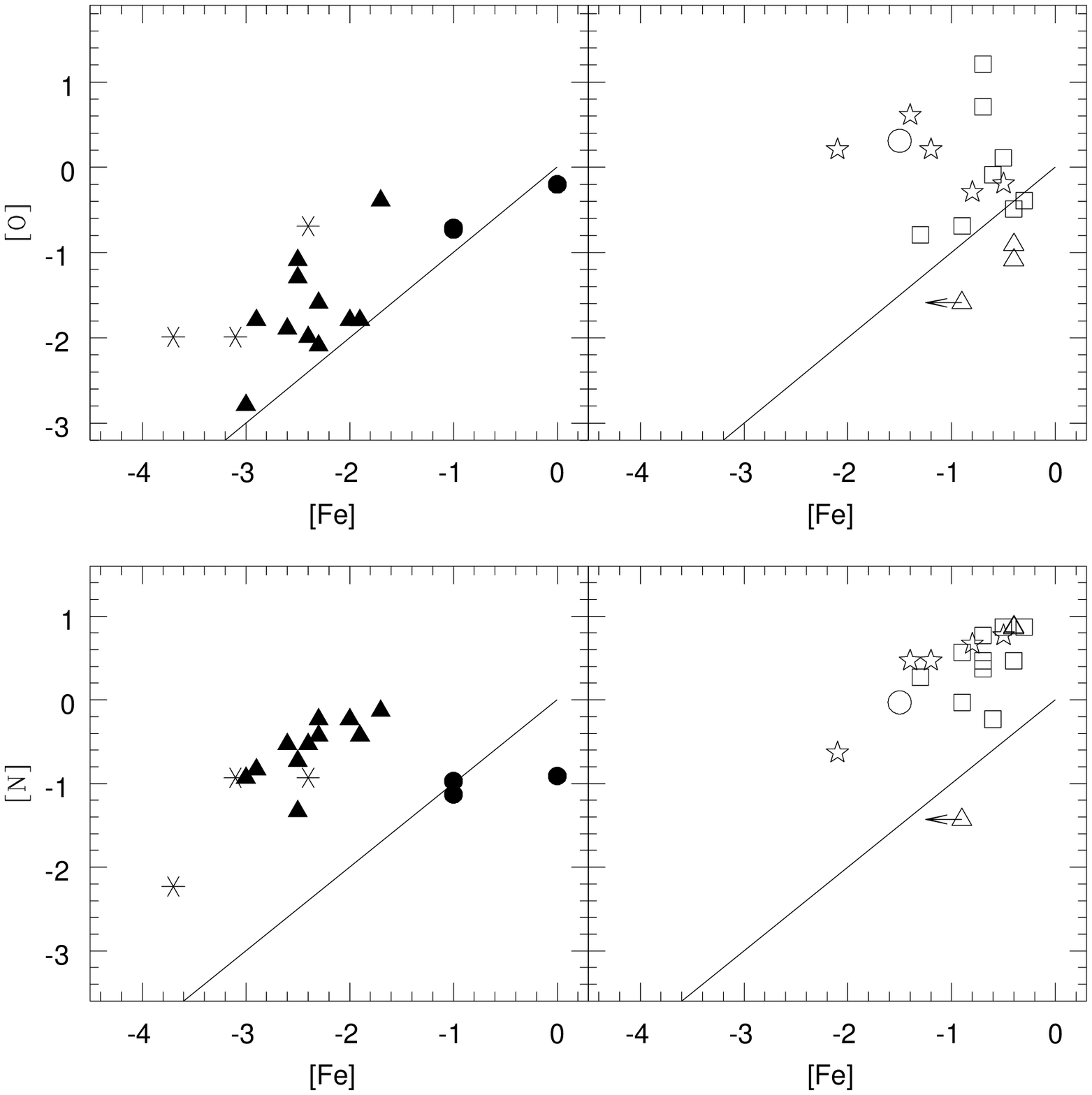}
\caption{[N] vs. [Fe] for RCBs (bottom left panel) and EHes (bottom right panel). 
[O] vs. [Fe] for RCBs (top left panel) and EHes (top right panel). The majority and minority class RCBs are
represented by filled triangles and asterisks, respectively. The hot and cool EHes are represented by open squares
and open stars, respectively. C-poor EHes are represented by open triangles. DY\,Cen is represented by open circle.
Filled circles denote the model predictions from \citet{crawford2020} for three Model IDs SOL8.57, SUB8.48, 
and SUB8.53, meeting the maximum criteria; details are in the text of Section 6.
[X]$=$[Fe] are denoted by solid lines, where X represents N, and O. [Fe],[X]$=$0,0 represents the Sun.}
\end{figure}

\begin{figure}
\epsscale{1.00}
\plotone{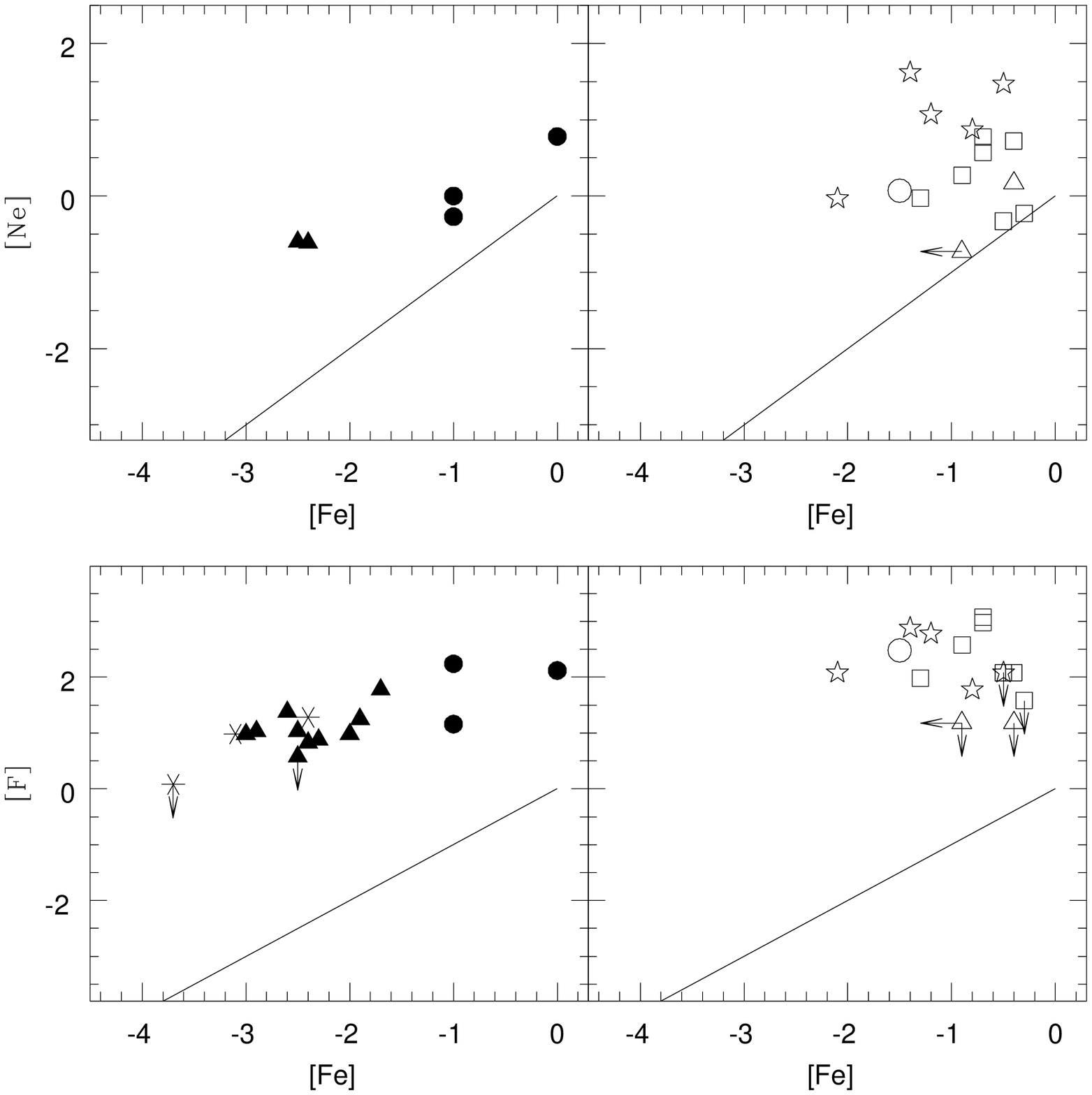}
\caption{[F] vs. [Fe] for RCBs (bottom left panel) and EHes (bottom right panel).
[Ne] vs. [Fe] for RCBs (top left panel) and EHes (top right panel). The majority and minority class RCBs are
represented by filled triangles and asterisks, respectively. The hot and cool EHes are represented by open squares
and open stars, respectively. C-poor EHes are represented by open triangles. DY\,Cen is represented by open circle.
Filled circles denote the model predictions from \citet{crawford2020} for three Model IDs SOL8.57, SUB8.48,
and SUB8.53, meeting the maximum criteria; details are in the text of Section 6.
[X]$=$[Fe] are denoted by solid lines, where X represents F, and Ne. [Fe],[X]$=$0,0 represents the Sun.}
\end{figure}

\begin{figure}
\epsscale{1.00}
\plotone{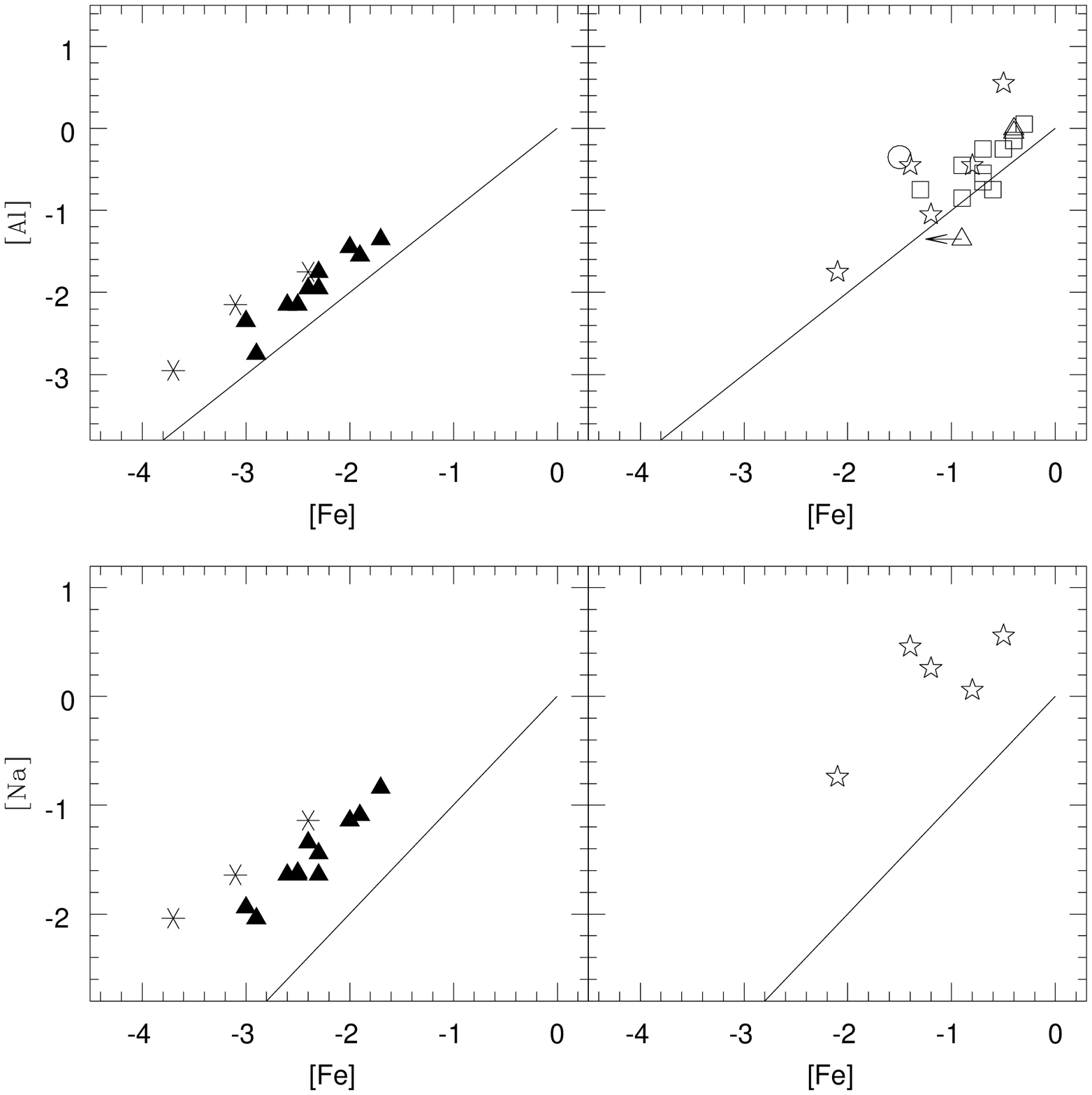}
\caption{[Na] vs. [Fe] for RCBs (bottom left panel) and EHes (bottom right panel).
[Al] vs. [Fe] for RCBs (top left panel) and EHes (top right panel). The majority and minority class RCBs are
represented by filled triangles and asterisks, respectively. The hot and cool EHes are represented by open squares
and open stars, respectively. C-poor EHes are represented by open triangles. DY\,Cen is represented by open circle.
[X]$=$[Fe] are denoted by solid lines, where X represents Na, and Al. [Fe],[X]$=$0,0 represents the Sun.}
\end{figure}

\begin{figure}
\epsscale{1.00}
\plotone{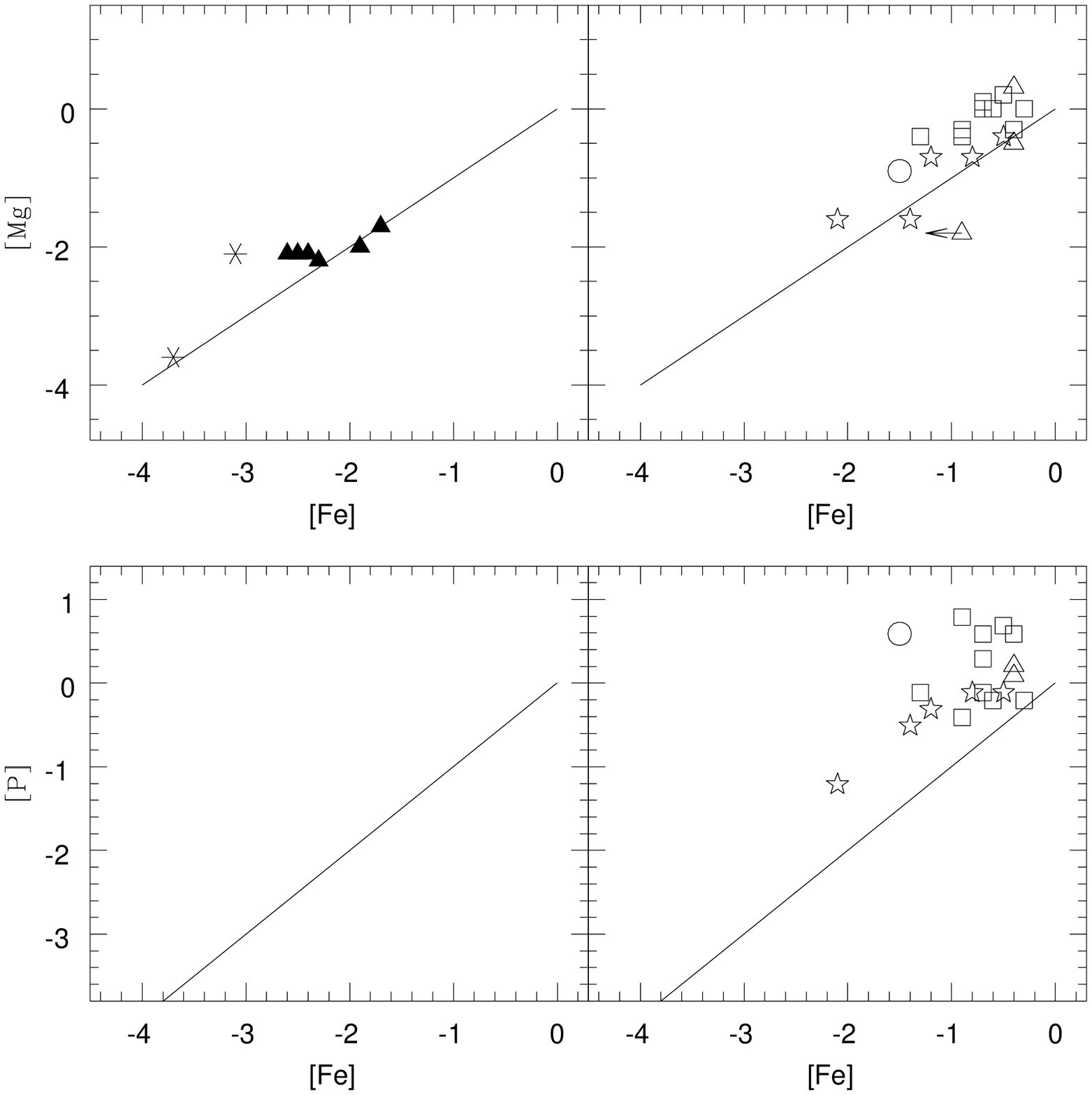}
\caption{[P] vs. [Fe] for RCBs (bottom left panel) and EHes (bottom right panel).
[Mg] vs. [Fe] for RCBs (top left panel) and EHes (top right panel). The majority and minority class RCBs are
represented by filled triangles and asterisks, respectively. The hot and cool EHes are represented by open squares
and open stars, respectively. C-poor EHes are represented by open triangles. DY\,Cen is represented by open circle.
[X]$=$[Fe] are denoted by solid lines, where X represents P, and Mg. [Fe],[X]$=$0,0 represents the Sun.}
\end{figure}

\begin{figure}
\epsscale{1.00}
\plotone{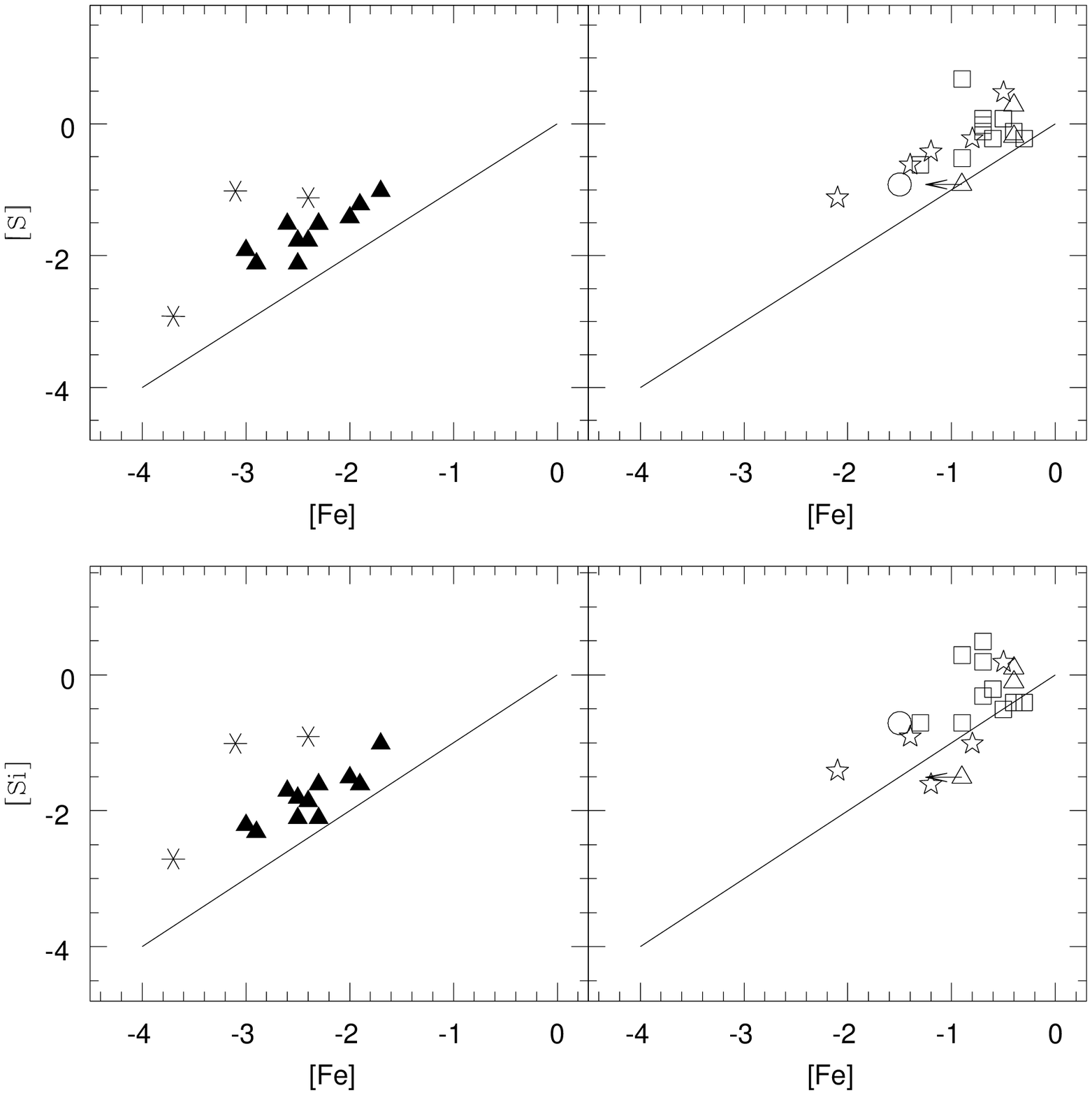}
\caption{[Si] vs. [Fe] for RCBs (bottom left panel) and EHes (bottom right panel).
[S] vs. [Fe] for RCBs (top left panel) and EHes (top right panel). The majority and minority class RCBs are
represented by filled triangles and asterisks, respectively. The hot and cool EHes are represented by open squares
and open stars, respectively. C-poor EHes are represented by open triangles. DY\,Cen is represented by open circle.
[X]$=$[Fe] are denoted by solid lines, where X represents Si, and S. [Fe],[X]$=$0,0 represents the Sun.}
\end{figure}

\begin{figure}
\epsscale{1.00}
\plotone{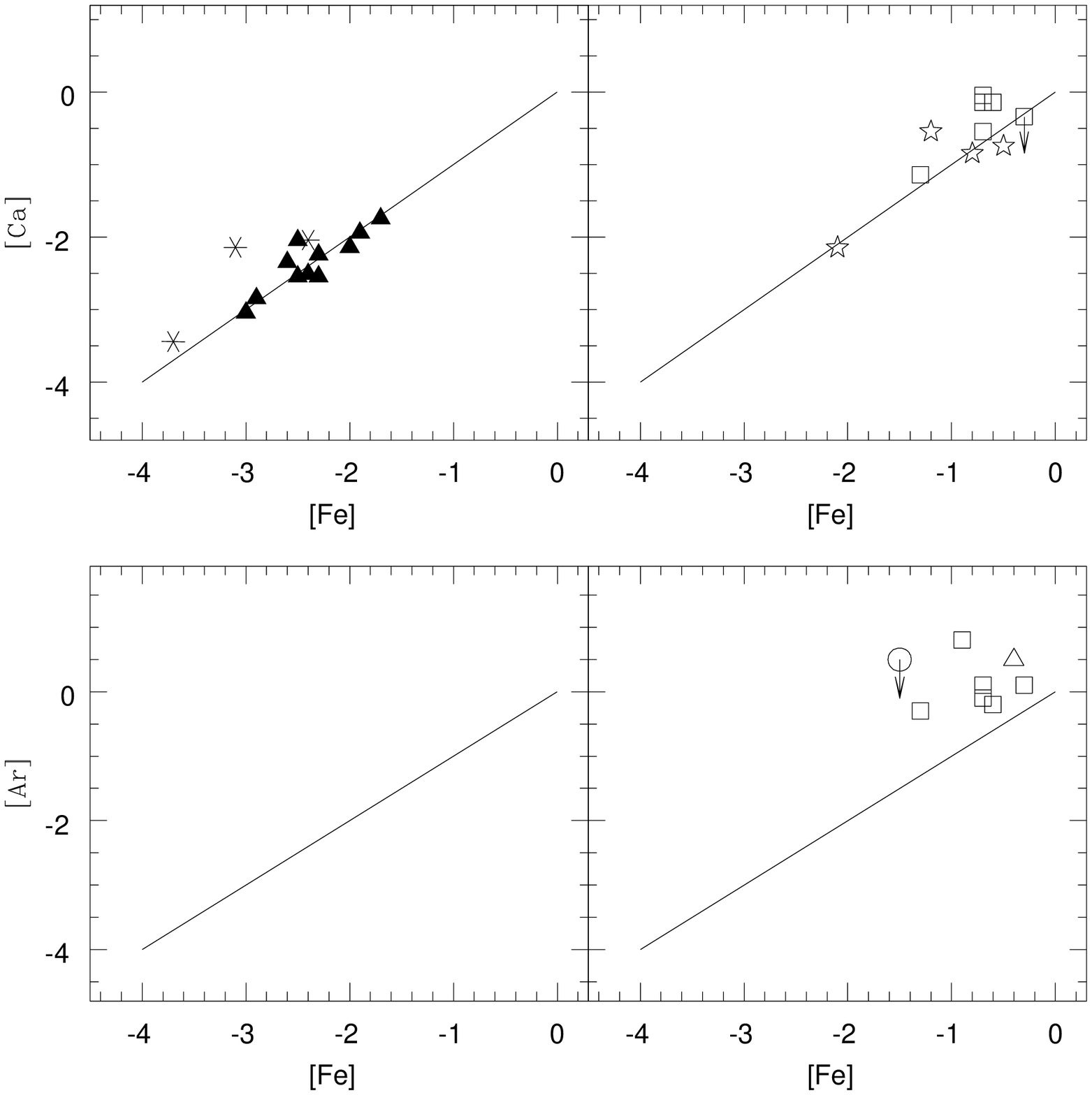}
\caption{[Ar] vs. [Fe] for RCBs (bottom left panel) and EHes (bottom right panel).
[Ca] vs. [Fe] for RCBs (top left panel) and EHes (top right panel). The majority and minority class RCBs are
represented by filled triangles and asterisks, respectively. The hot and cool EHes are represented by open squares
and open stars, respectively. C-poor EHes are represented by open triangles. DY\,Cen is represented by open circle.
[X]$=$[Fe] are denoted by solid lines, where X represents Ar, and Ca. [Fe],[X]$=$0,0 represents the Sun.}
\end{figure}

\begin{figure}
\epsscale{1.00}
\plotone{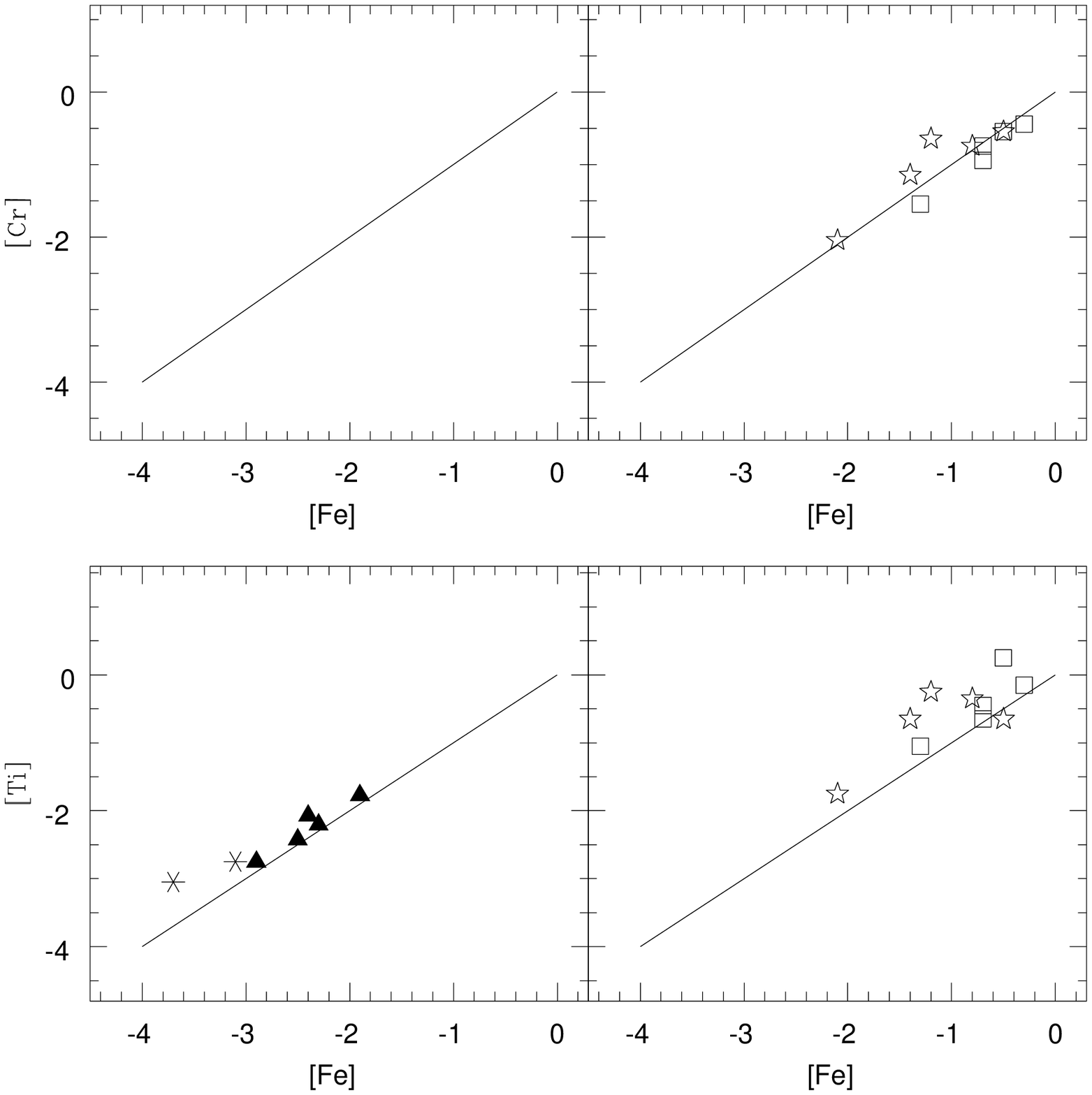}
\caption{[Ti] vs. [Fe] for RCBs (bottom left panel) and EHes (bottom right panel).
[Cr] vs. [Fe] for RCBs (top left panel) and EHes (top right panel). The majority and minority class RCBs are
represented by filled triangles and asterisks, respectively. The hot and cool EHes are represented by open squares
and open stars, respectively.
[X]$=$[Fe] are denoted by solid lines, where X represents Ti, and Cr. [Fe],[X]$=$0,0 represents the Sun.}
\end{figure}

\begin{figure}
\epsscale{1.00}
\plotone{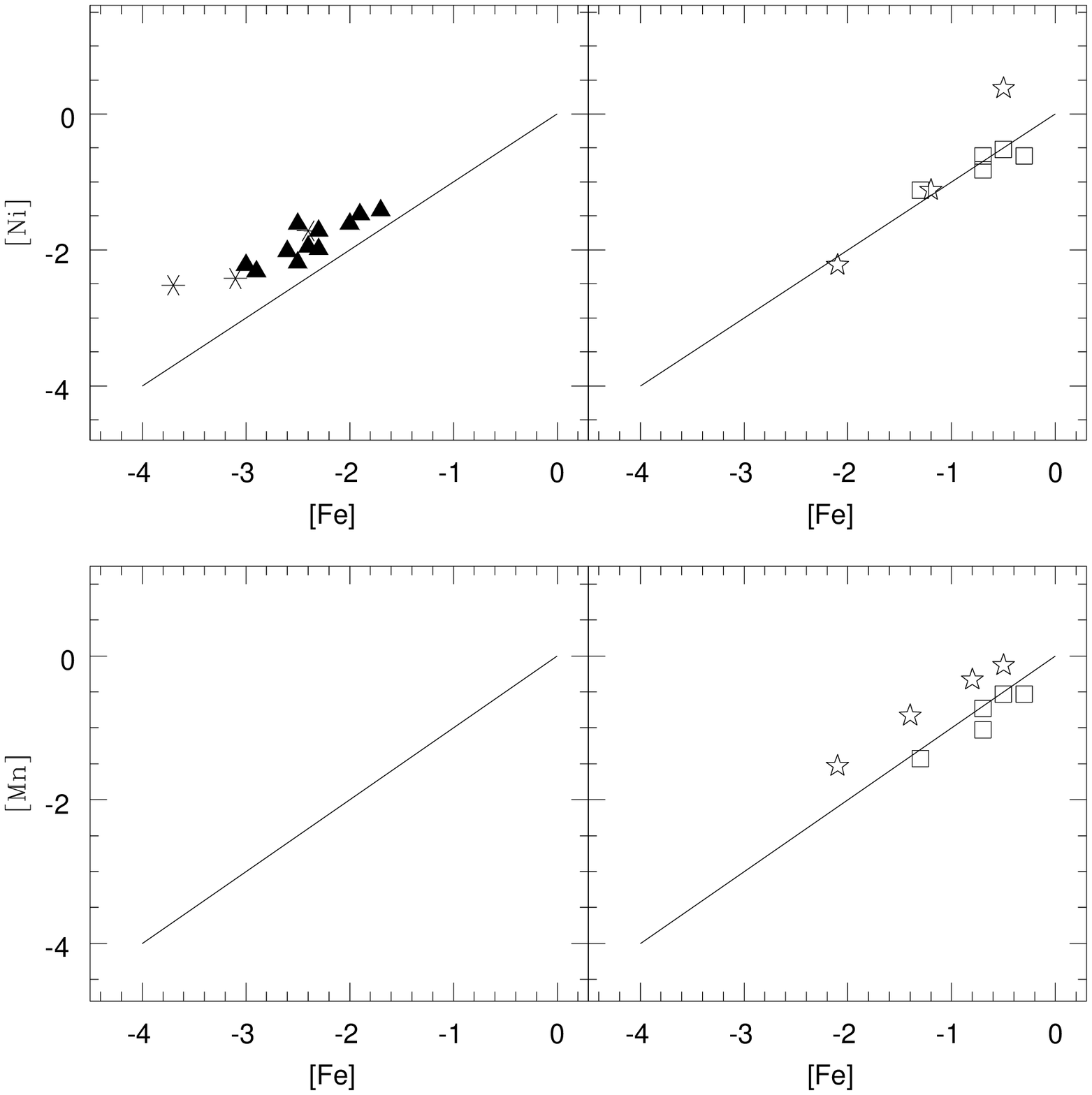}
\caption{[Mn] vs. [Fe] for RCBs (bottom left panel) and EHes (bottom right panel).
[Ni] vs. [Fe] for RCBs (top left panel) and EHes (top right panel). The majority and minority class RCBs are
represented by filled triangles and asterisks, respectively. The hot and cool EHes are represented by open squares
and open stars, respectively.
[X]$=$[Fe] are denoted by solid lines, where X represents Mn, and Ni. [Fe],[X]$=$0,0 represents the Sun.}
\end{figure}

\begin{figure}
\epsscale{1.00}
\plotone{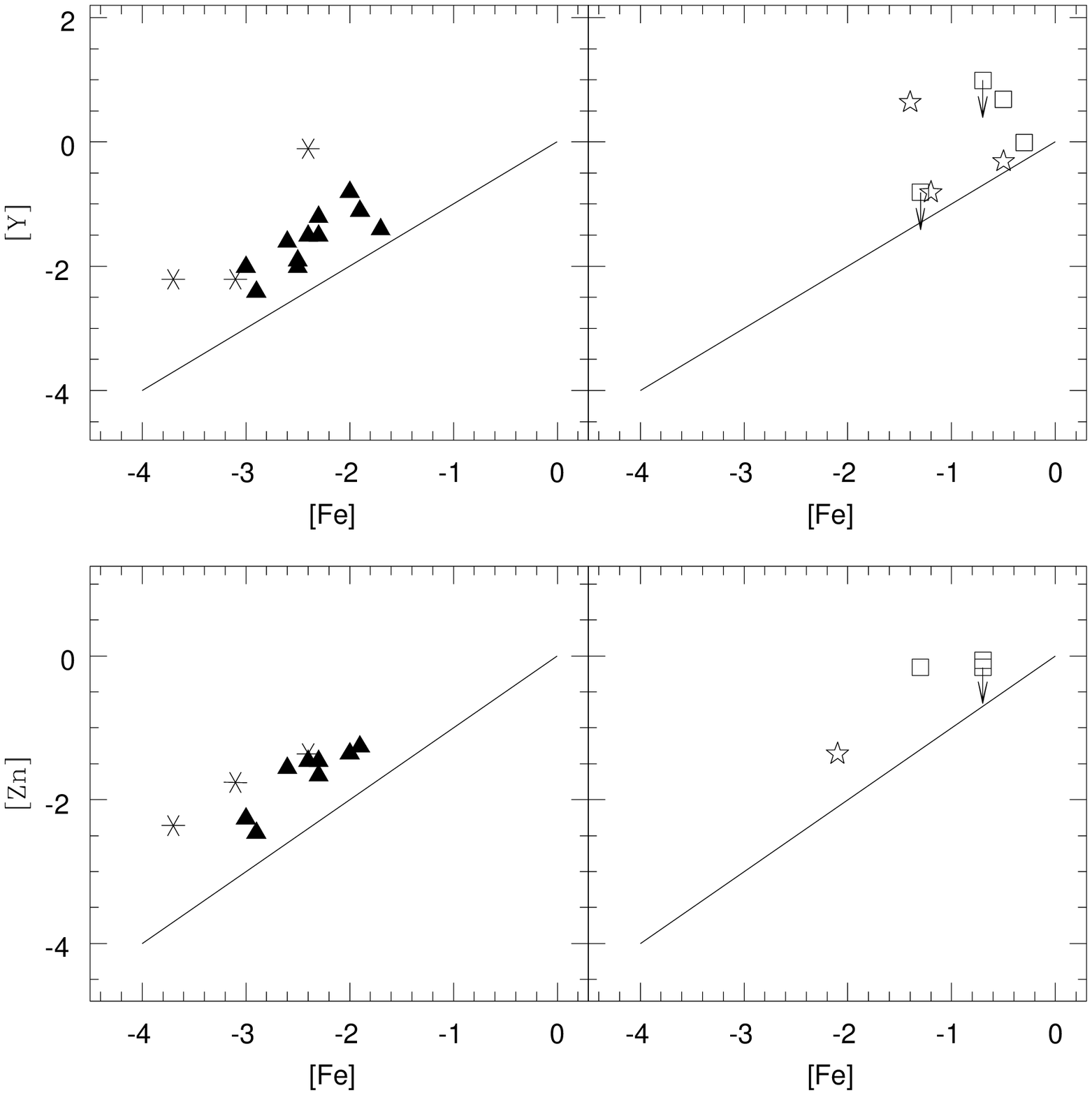}
\caption{[Zn] vs. [Fe] for RCBs (bottom left panel) and EHes (bottom right panel).
[Y] vs. [Fe] for RCBs (top left panel) and EHes (top right panel). The majority and minority class RCBs are
represented by filled triangles and asterisks, respectively. The hot and cool EHes are represented by open squares
and open stars, respectively.
[X]$=$[Fe] are denoted by solid lines, where X represents Zn, and Y. [Fe],[X]$=$0,0 represents the Sun.}
\end{figure}

\begin{figure}
\epsscale{1.00}
\plotone{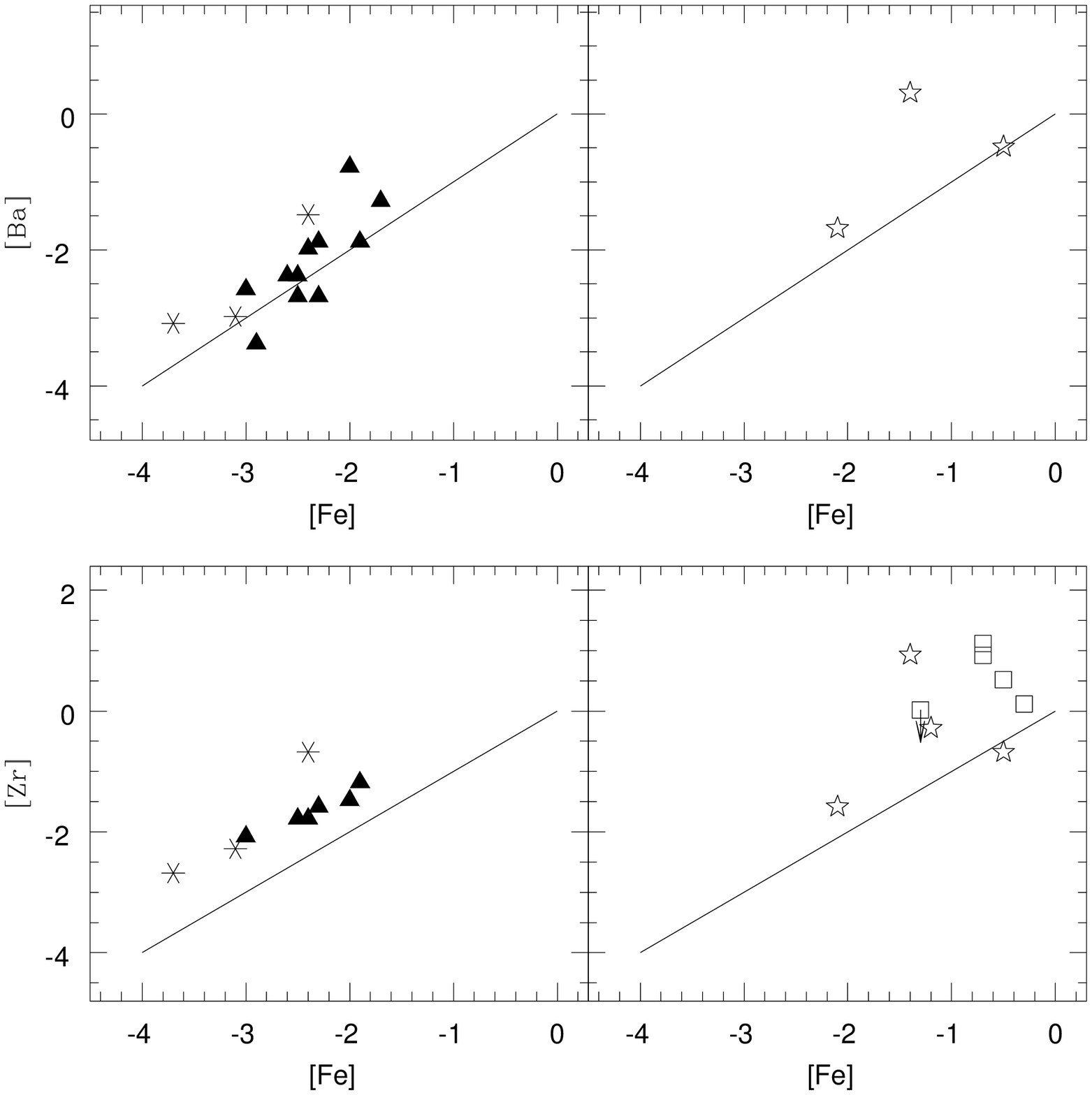}
\caption{[Zr] vs. [Fe] for RCBs (bottom left panel) and EHes (bottom right panel).
[Ba] vs. [Fe] for RCBs (top left panel) and EHes (top right panel). The majority and minority class RCBs are
represented by filled triangles and asterisks, respectively. The hot and cool EHes are represented by open squares
and open stars, respectively.
[X]$=$[Fe] are denoted by solid lines, where X represents Zr, and Ba. [Fe],[X]$=$0,0 represents the Sun.}
\end{figure}

\begin{figure}
\epsscale{1.00}
\plotone{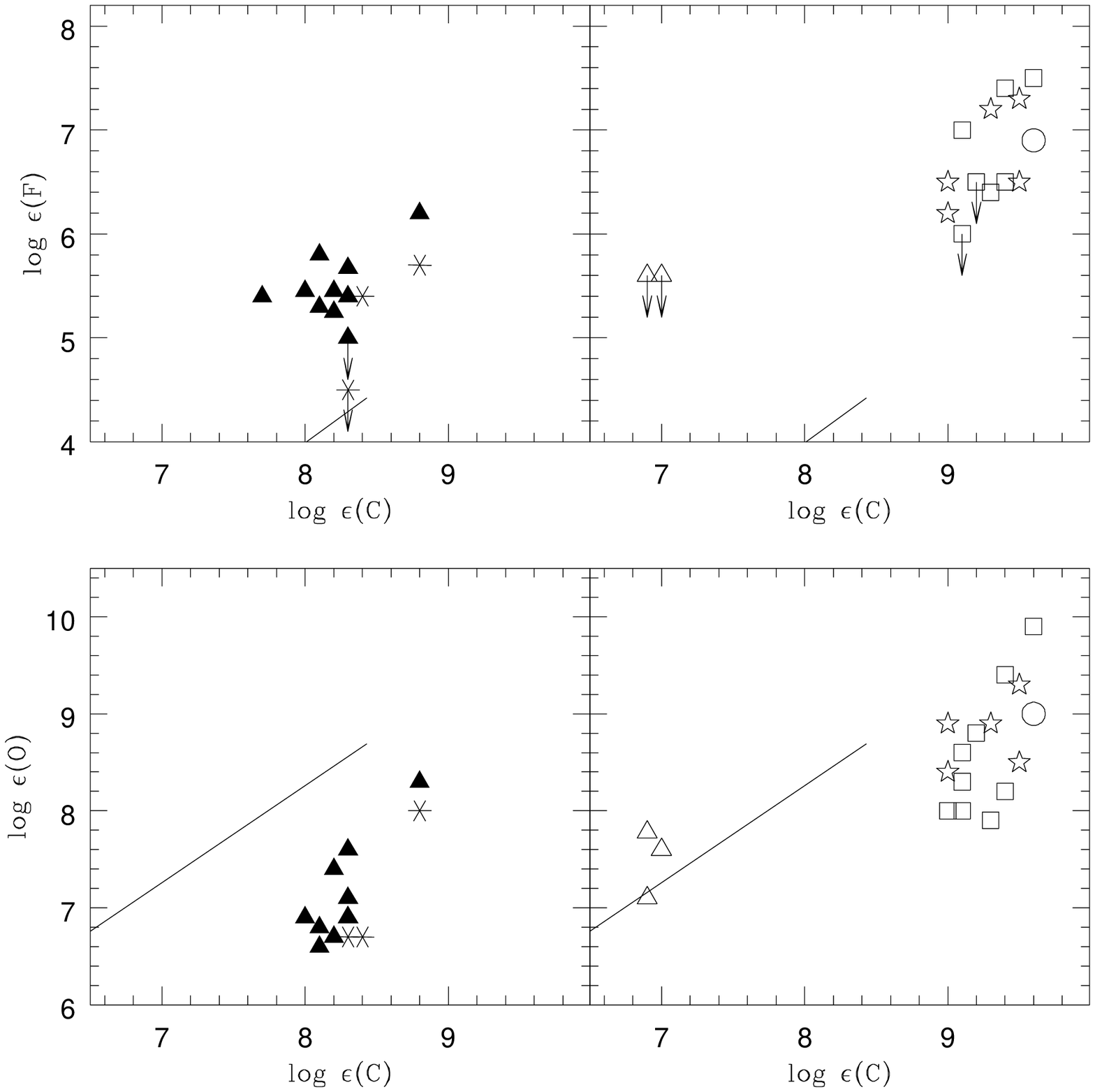}
\caption{$\log\epsilon$(O) vs. $\log\epsilon$(C) for RCBs (bottom left panel) and EHes (bottom right panel).
$\log\epsilon$(F) vs. $\log\epsilon$(C) for RCBs (top left panel) and EHes (top right panel). The majority and minority 
class RCBs are represented by filled triangles and asterisks, respectively. The hot and cool EHes are represented by 
open squares and open stars, respectively. C-poor EHes are represented by open triangles. DY\,Cen is represented by 
open circle. The solid lines denote the locus of the solar X/C ratios, where X represents O, and F.}
\end{figure}

\begin{figure}
\epsscale{1.00}
\plotone{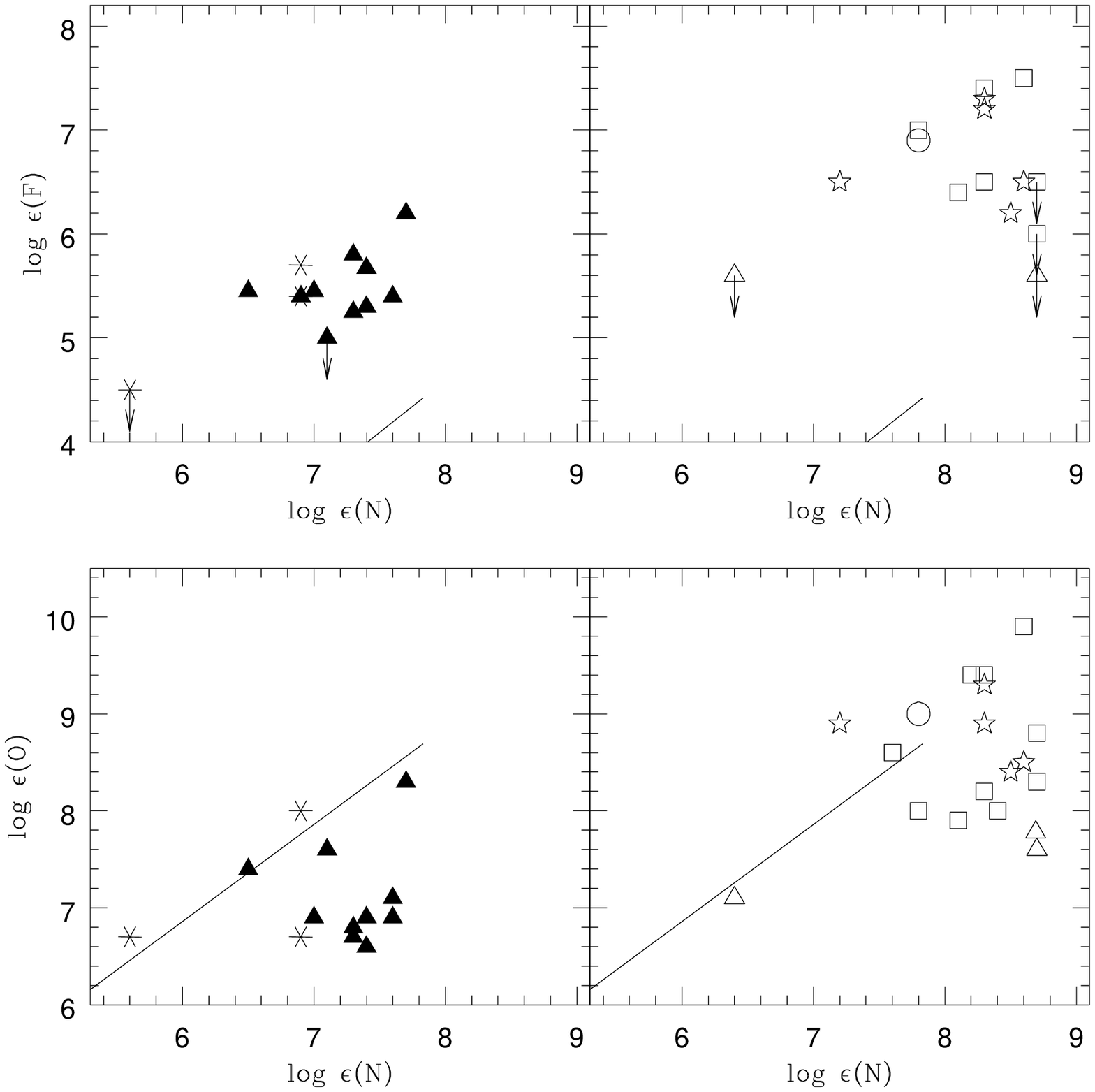}
\caption{$\log\epsilon$(O) vs. $\log\epsilon$(N) for RCBs (bottom left panel) and EHes (bottom right panel).
$\log\epsilon$(F) vs. $\log\epsilon$(N) for RCBs (top left panel) and EHes (top right panel). The majority and minority 
class RCBs are represented by filled triangles and asterisks, respectively. The hot and cool EHes are represented by 
open squares and open stars, respectively. C-poor EHes are represented by open triangles. DY\,Cen is represented by 
open circle. The solid lines denote the locus of the solar X/N ratios, where X represents O, and F.}
\end{figure}

\begin{figure}
\epsscale{1.00}
\plotone{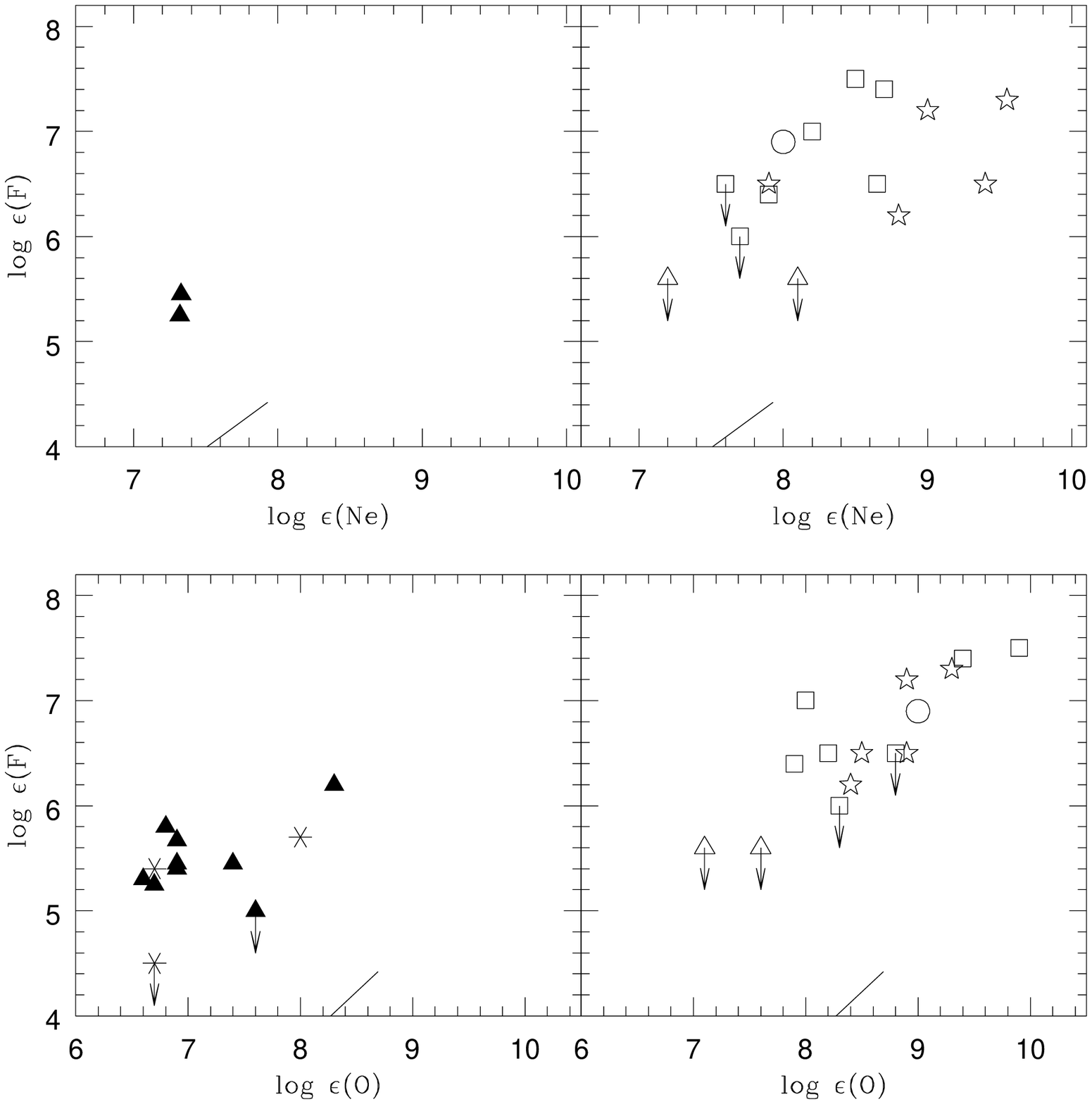}
\caption{$\log\epsilon$(F) vs. $\log\epsilon$(O) for RCBs (bottom left panel) and EHes (bottom right panel).
$\log\epsilon$(F) vs. $\log\epsilon$(Ne) for RCBs (top left panel) and EHes (top right panel). The majority and minority
class RCBs are represented by filled triangles and asterisks, respectively. The hot and cool EHes are represented by
open squares and open stars, respectively. C-poor EHes are represented by open triangles. DY\,Cen is represented by
open circle. The solid lines denote the locus of the solar F/X ratios, where X represents O, and Ne.}
\end{figure}

\clearpage
\thispagestyle{empty}
\begin{landscape} 
\begin{deluxetable}{lcccccccccccccccc}
\label{t:lines.uves}
\tabletypesize{\scriptsize}
\tablewidth{0pt}
\tablecolumns{17}
\tablecaption{Revised photospheric abundances of RCBs}
\tablehead{
\colhead{Element} & \colhead{GU\,Sgr} & \colhead{UX\,Ant} & \colhead{R\,CrB} &
\colhead{RS\,Tel} & \colhead{SU\,Tau} & \colhead{V482\,Cyg} & \colhead{FH\,Sct} &
\colhead{V2552\,Oph} & \colhead{V532\,Oph} & \colhead{RCB8\tablenotemark{a}} & \colhead{RCB10\tablenotemark{b}} &
\colhead{VZ\,Sgr} & \colhead{V\,CrA} & \colhead{V854\,Cen} & \colhead{FF\tablenotemark{c}} &
\colhead{Sun\tablenotemark{d}}}
\startdata
H   & \nodata     & 5.7       & 6.2     & 5.3       & 5.9     & 3.6         & 3.8         & 5.3         & 5.3         & 4.6         & 6.1     & 5.5       & 7.6         & 7.7         &  8.2         & 12.0  \\
Li  & $\leq -$0.3 & $\leq$1.3 & 2.1     & $\leq$0.2 & 1.1     & $\leq -$0.1 & $\leq -$0.8 & $\leq -$0.4 & $\leq -$0.3 & $\leq -$0.2 & \nodata & $\leq$0.8 & $\leq -$0.2 & $\leq -$0.2 &  3.4         &  1.05 \\
C   & 8.1         & 8.3       & 8.8     & 8.3       & 8.0     & 8.3         & 7.7         & 8.1         & 8.2         & 8.3         & 8.2     & 8.8       & 8.4         & 8.3         &  9.7         &  8.43 \\
$\Delta$C\tablenotemark{e} & 1.4 & 1.2 & 0.7 & 1.2 &  1.5 &  1.2 & 1.8 & 1.4 & 1.3 &  1.2 & 1.3 & 0.7 &  1.1 & 1.2 & 0.8 & \nodata \\
N   & 7.3         & 7.1       & 7.7     & 7.6       & 7.0     & 7.6         & 6.9         & 7.4         & 7.3         & 7.4         & 6.5     & 6.9       & 6.9         & 5.6         &  8.1         &  7.83 \\
O   & 6.8         & 7.6       & 8.3     & 7.1       & 6.9     & 6.9         & 5.9         & 6.6         & 6.7         & 6.9         & 7.4     & 8.0       & 6.7         & 6.7         &  8.6         &  8.69 \\
F   & 5.8         & 5.0       & 6.2     & \nodata   & 5.5     & 5.4         & 5.4         & 5.3         & 5.3         & 5.7         & 5.5     & 5.7       & 5.4         & 4.5         &  $<$5.6      &  4.42 \\
Ne  & \nodata     & \nodata   & \nodata & \nodata   & \nodata & \nodata     & \nodata     & \nodata     & 7.3         & \nodata     & 7.3     & \nodata   & \nodata     & \nodata     &  \nodata     &  7.93 \\
Na  & 4.6         & 4.6       & 5.4     & 4.8       & 4.2     & 5.1         & 4.3         & 4.6         & 4.9         & 5.2         & 4.6     & 5.1       & 4.6         & 4.2         &  6.0         &  6.24 \\
Mg  & 5.5         & \nodata   & 5.9     & \nodata   & \nodata & \nodata     & \nodata     & 5.4         & 5.5         & 5.6         & 5.5     & \nodata   & 5.5         & 4.0         &  5.7         &  7.60 \\
Al  & 4.3         & \nodata   & 5.1     & 4.7       & 3.7     & 5.0         & 4.1         & 4.5         & 4.5         & 4.9         & 4.3     & 4.7       & 4.3         & 3.5         &  5.5         &  6.45 \\
Si  & 5.8         & 5.7       & 6.5     & 5.9       & 5.2     & 6.0         & 5.3         & 5.4         & 5.7         & 5.9         & 5.4     & 6.6       & 6.5         & 4.8         &  6.7         &  7.51 \\
P   & \nodata     & \nodata   & \nodata & \nodata   & \nodata & \nodata     & \nodata     & \nodata     & \nodata     & \nodata     & \nodata & \nodata   & \nodata     & \nodata     &  \nodata     &  5.41 \\
S   & 5.6         & 5.0       & 6.1     & 5.6       & 5.0     & 5.7         & 5.2         & 5.6         & 5.4         & 5.9         & 5.4     & 6.0       & 6.1         & 4.2         &  6.1         &  7.12 \\
Ca  & 4.0         & 4.3       & 4.6     & 4.1       & 3.5     & 4.2         & 3.3         & 3.8         & 3.8         & 4.4         & 3.8     & 4.3       & 4.2         & 2.9         &  4.7         &  6.34 \\
Ti  & \nodata     & \nodata   & \nodata & \nodata   & 2.2     & \nodata     & \nodata     & 2.8         & 2.9         & 3.2         & 2.5     & \nodata   & 2.2         & 1.9         &  3.8         &  4.95 \\
Fe  & 4.9         & 5.0       & 5.8     & 5.2       & 4.6     & 5.5         & 4.5         & 5.2         & 5.1         & 5.6         & 5.0     & 5.1       & 4.4         & 3.8         &  5.8         &  7.50 \\
Ni  & 4.2         & 4.6       & 4.8     & 4.5       & 3.9     & 4.6         & 4.0         & 4.2         & 4.3         & 4.7         & 4.0     & 4.5       & 3.8         & 3.7         &  5.4         &  6.22 \\
Zn  & 3.0         & \nodata   & \nodata & 3.1       & 2.1     & 3.2         & 2.3         & 2.9         & 3.1         & 3.3         & \nodata & 3.2       & 2.8         & 2.2         &  4.6         &  4.56 \\
Y   & 0.6         & 0.3       & 0.8     & 0.7       & $-$0.2  & 1.4         & 0.2         & 1.0         & 0.7         & 1.1         & 0.2     & 2.1       & \nodata     & \nodata     &  3.4         &  2.21 \\
Zr  & \nodata     & \nodata   & \nodata & \nodata   & \nodata & 1.1         & 0.5         & 1.0         & 0.8         & 1.4         & 0.8     & 1.9       & 0.3         & $-$0.1      &  2.7         &  2.58 \\
Ba  & $-$0.2      & $-$0.2    & 0.9     & 0.3       & $-$1.2  & 1.4         & $-$0.4      & $-$0.5      & 0.2         & 0.3         & $-$0.5  & 0.7       & $-$0.8      & $-$0.9      &  1.1         &  2.18 \\
\enddata
\tablenotetext{a}{ASAS-RCB-8}
\tablenotetext{b}{ASAS-RCB-10}
\tablenotetext{c}{V4334\,Sgr (Sakurai's object), a FF product, shows no detectable neutral fluorine lines \citep{pandeyetal2008}}
\tablenotetext{d}{\citet{asplund09}}
\tablenotetext{e}{$\Delta$C=C(assumed for the model)$-$C(derived from C$_{2}$ bands). For the RCB stars assumed C/He=1\% or C=9.5 dex, and for FF object assumed C/He=10\% or C$=$10.5 dex.}
\end{deluxetable}
\end{landscape}


\end{document}